\documentclass[preprint]{aastex}
\usepackage{natbib}
\received{}
\accepted{}

\author{
David E. Johnston,\altaffilmark{1,2}
Gordon T. Richards,\altaffilmark{3}
Joshua A. Frieman,\altaffilmark{1,2,4}
Charles R. Keeton,\altaffilmark{1,5}
Michael A. Strauss,\altaffilmark{3}
Robert H. Becker,\altaffilmark{6,7}
Richard L. White ,\altaffilmark{8}
Eric T. Johnson,\altaffilmark{2}
Zhaoming Ma,\altaffilmark{1}
Mark SubbaRao,\altaffilmark{1,9}
Neta A. Bahcall ,\altaffilmark{3}
Mariangela Bernardi ,\altaffilmark{18}
Jon Brinkmann, \altaffilmark{10}
Daniel J. Eisenstein, \altaffilmark{11}
Masataka Fukugita,  \altaffilmark{12}
Patrick B. Hall,\altaffilmark{3,13}
Naohisa Inada,\altaffilmark{14}
Gillian R. Knapp,\altaffilmark{3}
Bartosz Pindor,\altaffilmark{3}
David J. Schlegel \altaffilmark{3}
Ryan Scranton,\altaffilmark{15}
Erin S. Sheldon,\altaffilmark{1,2}
Donald P. Schneider , \altaffilmark{16}
Alexander S. Szalay, \altaffilmark{17}
and Donald G. York \altaffilmark{1}
}
\altaffiltext{1}{Department of Astronomy and Astrophysics, The University of
Chicago, 5640 South Ellis Avenue, Chicago, IL 60637.}
\altaffiltext{2}{Center for Cosmological Physics, The University of 
Chicago, 5640 South Ellis Avenue, Chicago, IL 60637.}
\altaffiltext{3}{Princeton University Observatory, Peyton Hall, Princeton, NJ 08544.}
\altaffiltext{4}{Fermi National Accelerator Laboratory, P.O. Box 500, Batavia,
IL 60510.}
\altaffiltext{5}{Hubble Fellow.}
\altaffiltext{6}{Physics Department, University of California, Davis, CA 95616.}
\altaffiltext{7}{IGPP-LLNL, L-413, 7000 East Avenue, Livermore, CA 94550.}
\altaffiltext{8}{Space Telescope Science Institute, 3700 San Martin Drive, Baltimore, MD 21218.}
\altaffiltext{9}{Adler Planetarium \& Astronomy Museum, Chicago, IL 60605.}
\altaffiltext{10}{Apache Point Observatory, 2001 Apache Point Road, P.O Box 59,Sunspot, NM, 88349}
\altaffiltext{11}{Steward Observatory, University of Arizona, 
933 North Cherry Avenue, Tucson, AZ 85721.}
\altaffiltext{12}{Institute for Cosmic Ray Research, 
University of Tokyo, 5-1-5 Kashiwa, Kashiwa City, Chiba 277-8582.}
\altaffiltext{13}{Departamento de Astronomia y Astrofisica, Facultad de Fisica,
Pontificia Universidad Catolica de Chile, Casilla 306, Santiago 22, Chile.}
\altaffiltext{14}{Department of Physics, University of Tokyo, 
Hongo 7-3-1, Bunkyo-ku, Tokyo, 113-0033, Japan.}
\altaffiltext{15}{Department of Physics \& Astronomy, University of 
Pittsburgh, 3941 O'Hara St., Pittsburgh, PA 15260.}
\altaffiltext{16}{Department of Astronomy and Astrophysics, The 
 Pennsylvania State University, 525 Davey Laboratory, University Park, PA 16802}
\altaffiltext{17}{Department of Physics and Astronomy, The Johns Hopkins University,
Baltimore, MD 21218}
\altaffiltext{18}{Department of Physics, Carnegie Mellon University, Pittsburgh, PA 15213}

\begin{document}

\title{SDSS~J0903+5028: A New Gravitational Lens}

\begin{abstract}
We report the discovery of a new gravitationally lensed quasar 
from the Sloan Digital Sky Survey, SDSS~J090334.92+502819.2. 
This object was targeted for SDSS spectroscopy 
as a Luminous Red Galaxy (LRG), but
manual examination of the spectrum showed the presence of
a quasar at $z\simeq 3.6$ in addition to a red galaxy at $z=0.388$, 
and the SDSS image showed a second possible QSO image nearby.
Follow-up
imaging and spectroscopy confirmed the lensing hypothesis. 
In images taken at the ARC 3.5-meter telescope, two 
quasars are separated by $2\farcs8$; the lensing
galaxy is clearly seen and is blended with one of
the quasar images. Spectroscopy taken at the Keck II telescope
shows that the quasars have identical redshifts of
$z\simeq 3.6$ and both show the presence of the same broad 
absorption line-like troughs. We present simple lens models which account 
for the geometry and magnifications. The lens galaxy lies near
two groups of galaxies and may be a part of them. The models 
suggest that the groups may contribute considerable shear and 
may have a strong effect on the lens configuration. 
\end{abstract}


\section{Introduction}

Gravitational lenses have become important astrophysical and
cosmological tools in several ways. The frequency of 
lensing is in principle sensitive to the dark energy density  
(Fukugita et al. 1990, Turner 1990, Fukugita \& Turner 1991, Kochanek 1995, 
but see Keeton 2002) 
and the matter density of the Universe \citep{mort}. Lens 
statistics also probe the properties of the 
lensing galaxy systems, such as their 
mass distribution, potential well depth and extinction
\citep{chen,chae,kkf,malhot,keeton1,keetmad}. In addition,   
measurement of the time delay between images in individual 
lensed quasars can be used to measure the Hubble parameter \citep{refs}.

Since the discovery of the first double quasar Q0957+561
\citep{wcw,schild}, about 80 lensed quasars have been 
discovered\footnote{see http://cfa-www.harvard.edu/castles/}.
In this paper, we report the discovery of another lensed
quasar, SDSS~J090334.92+502819.2 (hereafter SDSS~J0903+5028), 
discovered in the Sloan Digital Sky Survey data (SDSS; \citep{yaa00}). 
There have been several gravitational lenses discovered previously in the 
SDSS data \citep{inada1,inada2,burles,pin,morgan}, 
but this one is unusual in the way it was found. The standard algorithm for 
selecting lens candidates in the SDSS involves 
looking for deviations from PSF profiles
for spectroscopically confirmed quasars \citep{inada1,pin02}. By contrast, 
SDSS~J0903+5028 was selected for follow-up based on 
the presence of $z\simeq 3.6$ quasar features superimposed on  
the SDSS fiber spectrum of a $z=0.388$ luminous red galaxy (LRG). 
Moreover, the SDSS image 
showed another source about $2\farcs5$ from the spectroscopically 
targeted galaxy as well as the presence of surrounding 
galaxies with colors similar to the LRG. While this system 
may be unusual, it is not unprecedented: the well-known 
lens 2237+0305 \citep{huchra} was discovered serendipitously 
in a galaxy redshift survey, and such cases are 
expected \citep{kochred,mortweb2}.

Follow-up 
spectroscopy at the ARC 3.5-meter telescope showed that 
both main image components contain flux from a $z\simeq 3.6$
quasar with strong BAL-like associated absorption \citep{foltz}, 
strongly suggesting that this is a lensed 
system. Subsequent $r$ and $i$ band imaging at the ARC 3.5-meter 
telescope in better seeing revealed the lens geometry 
more clearly, as shown in Figure~\ref{fig:spicam-zoom},
and enabled us to model the lens system.  
Finally, higher signal-to-noise ratio, higher-resolution spectra 
were taken at the Keck II telescope. Taken together, these 
data make a firm case that SDSS~J0903+5028 comprises two 
images of a high-redshift quasar lensed by a massive, red, 
foreground galaxy in a group.

This paper is organized as follows. 
In Section 2 we describe the SDSS data on this object, discuss  
how it was selected for follow-up, and 
describe the spectroscopic and imaging data
from the ARC 3.5m telescope and spectroscopic data from the 
Keck telescope. In Section 3, we 
fit a simple two-quasar + galaxy model to the ARC 3.5m images 
and extract positions and magnitudes for the three components. 
With this information, we fit a lens model,
estimate the velocity dispersion of the lens galaxy, and 
study the quadrupole moment of the lensing potential.
We also decompose the SDSS and Keck 
spectra into quasar and galaxy components and find flux ratios 
consistent with the imaging data. We conclude in Section 4.

\section{Observations}

The data on SDSS~J0903+5028 consist of the following. The
object was observed in routine SDSS imaging
in January 2001. Based on its colors and 
brightness, it was targeted for SDSS spectroscopy 
as a luminous red galaxy (LRG) and spectroscopic observations
were taken in March 2001. 
Due to the presence of both quasar and galaxy features in the 
SDSS spectrum, the object 
was included in a list of promising lens candidates for 
follow-up observations. Spectroscopy of the two main components 
on the ARC 3.5m telescope in October 2002 
revealed that they both contain very similar quasar  
features in addition to galaxy spectral features. This observation was 
followed by higher 
quality spectroscopy with the Keck II telescope and deeper, 
better seeing-quality imaging data with the ARC 3.5m telescope. 
In the following, we describe each of these data 
sets in detail.

\subsection{SDSS Data} 

The Sloan Digital Sky Survey is a wide-field photometric and spectroscopic 
survey being carried out by the Astrophysical Research Consortium (ARC) 
at the Apache Point Observatory in New Mexico \citep{yaa00}. The  
SDSS multi-CCD camera \citep{gcr98} will produce images for $\sim 5 \times 10^8$ 
objects in five optical bands $u, g, r, i, z$ 
to a detection limit of approximately $r = 22.2$. 
The photometric pipeline software is described in Lupton et al. (2001).
The photometric calibration is described in Fukugita et al. (1996) and 
Smith et al. (2002) 
and the astrometric calibration in Pier et al. (2003). 
Galaxies, quasars, 
stars, and other sources identified in SDSS imaging 
are targeted for SDSS spectroscopy based 
on several selection criteria \citep{Stoughton}; for a description of the
selection algorithm for luminous red galaxies (LRGs), see 
Eisenstein et al. (2001). 

\begin{figure}[p]
\epsscale{1.0}
\plotone{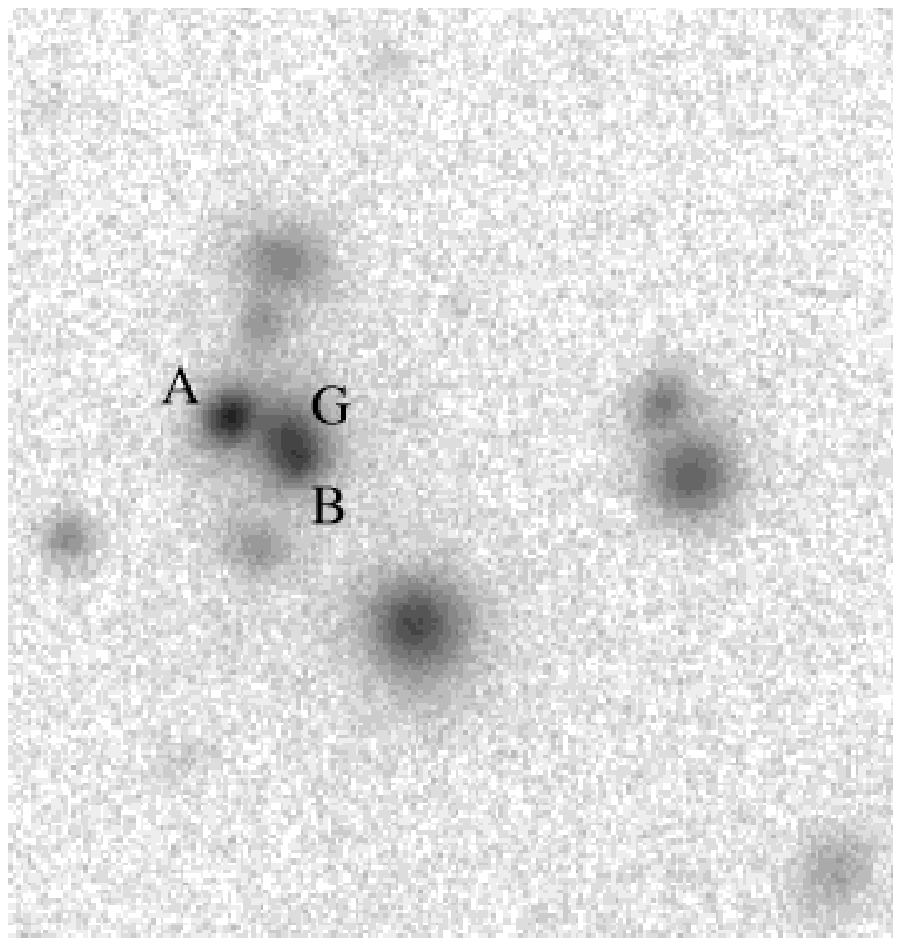}
\caption{SPIcam $r$-band image of area around SDSS~J0903+5028. North 
is up and East is to the left. The scale of the image is $35\arcsec$ across,
the pixel scale is 0\farcs14/pixel and the seeing is 1\farcs1.
The objects labeled A and B are the quasar images; the galaxy
is labeled G and is blended with quasar B. These other galaxies may be
a small group or part of two nearby clusters. 
\label{fig:spicam-zoom}}
\end{figure}

SDSS~J0903+5028 was imaged on 2001 January 26 with SDSS 
identifiers: run 2074, camera column 2, field 113. This field is included 
in the recently released SDSS Data Release 1 
\footnote{also see http://www.sdss.org/dr1/} 
~\cite{abazajian}. 
SDSS~J0903+5028 is the close pair of 
objects near the center of the $r$ band image shown in
Figure~\ref{fig:spicam-zoom}. The Western 
member of the pair is a blend of galaxy and quasar light 
and corresponds to the combined object(s) labeled  
G, B in Figure~\ref{fig:spicam-zoom}. 
In the photometric reduction 
used for spectroscopic targeting (rerun 0), this object 
was assigned SDSS identification number id 229 in this field. 
In the current `best' reduction (rerun 21), using 
an improved version of the photometric pipeline,  
the object has id 186. The Eastern member of the 
pair---component A in Figure~\ref{fig:spicam-zoom}---was also identified 
by the photometric pipeline (id 228 in rerun 0 and id 185 in 
rerun 21, with the same run, camcol, and field numbers as above).  
It was identified as a faint galaxy, not a point source, 
most likely because of its close proximity to the B/G component.

In Table~\ref{tab-image-sdssmag} we provide SDSS photometric 
\citep{hogg,smith} and 
astrometric \citep{pier}
parameters for both components from the `best' rerun 21. 
Since   
the seeing measured in bands $u,g,r,i,z$ was $1\farcs83$,
$1\farcs73$, $1\farcs50$, $1\farcs40$, $1\farcs50$, respectively,  
comparable to the separation of the two images, the A and G/B components 
were not fully deblended from each other by the SDSS photometric pipeline. 
As a result, one should be cautious 
about interpreting the SDSS photometric parameters for this object.
More accurate astrometric and photometric information for the 
different components, based on subsequent ARC 3.5m imaging in better 
conditions, is given in Table~\ref{tab-image-mod}.  

\begin{figure}[p]
\epsscale{1.0}
\plotone{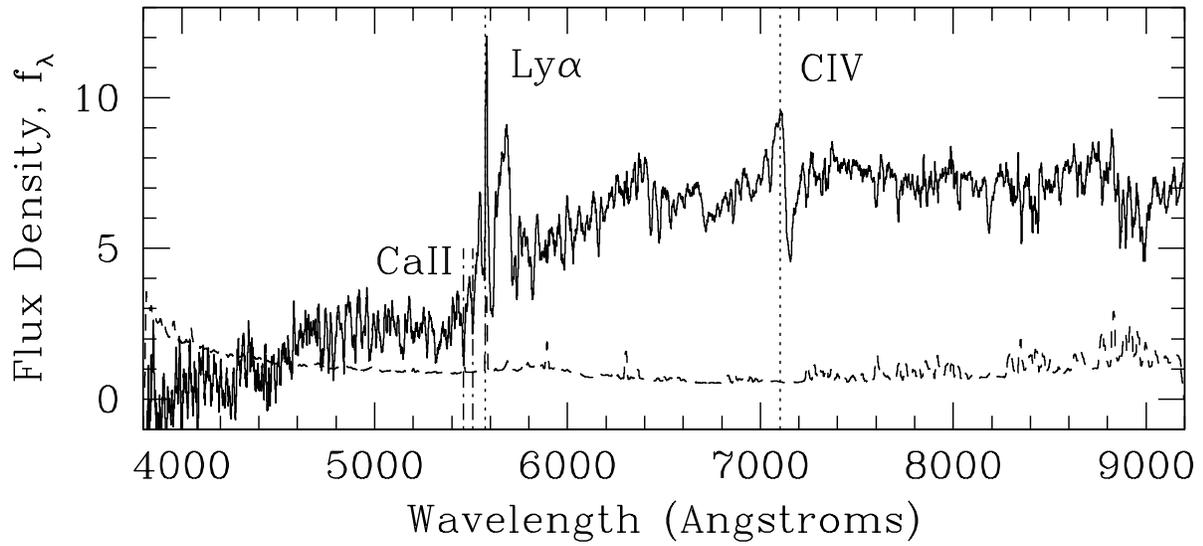}
\caption{SDSS spectrum of SDSS~J0903+5028 (smoothed by 9 pixels).  
The error spectrum (also smoothed by 9 pixels) is given by the 
dashed line. Dotted lines mark the centers of Ly$\alpha$ and 
CIV emission for $z=3.584$. The flux units are
$10^{-17}$ ergs/s/cm$^{2}$/\AA.
\label{fig:fig1}}
\end{figure}

The G/B component was targeted for SDSS spectroscopy as
a `cut II' luminous red galaxy (LRG) (Eisenstein et al. 2001;  
see Blanton, et al. 2003 for a description of the 
spectroscopic tiling algorithm) and was 
likely boosted above the flux limit of 
the LRG sample by the addition of the blended quasar light.  
Its spectrum was taken 
on 2001 March 24 with SDSS identifiers: Plate 552,
Fiber 221, MJD 51992. This plate was observed for 
$8\times 15$-minute exposures, yielding combined spectra of somewhat higher  
S/N than is typical for the survey. 
The $3\arcsec$ spectroscopic fiber was centered on 
$09^{\rm h} 03^{\rm m} 34^{\rm s}.92+50^{\circ} 28\arcmin 19\farcs2$ (J2000); 
in Figure~\ref{fig:spicam-zoom}, this corresponds approximately 
to centering on the G component. The SDSS spectrum, shown in 
Figure~\ref{fig:fig1}, clearly shows absorption features of an early-type 
galaxy at redshift $z=0.388$ (e.g., the Ca H and K lines at 
5463 and 5510 \AA), along with strong quasar emission 
lines with a peak \ion{C}{4} redshift of $z=3.584$. 

The surrounding field in Figure~\ref{fig:spicam-zoom} 
shows several fainter galaxies with colors 
similar to those of the G/B component of SDSS~J0903+5028, 
suggesting the presence of a small galaxy group associated with the LRG. 
We have applied a group finding algorithm which looks for a red-sequence 
in color-magnitude space for over-dense regions, the maxBCG algorithm   
\cite{annis,bahcall}. The algorithm does not find a cluster or group
centered at the lens galaxy since it is not a local maximum of the
galaxy density on cluster length scales. However it does find two small nearby
clusters, both
about $5\farcm7$ away and both with photometric redshifts of
z=$0.44 \pm 0.03$. The lens galaxy sits right between these two
clusters and they are aligned north and south of the lens. 
The properties of these photometric clusters are summarized
in Table~\ref{tab-maxbcg}. The actual physics of these two
clusters will require a detailed spectroscopic study:
they may be part of one bigger cluster or set of merging clusters
that span the entire region including
the compact group of galaxies surrounding the lens galaxy 
shown in Figure~\ref{fig:spicam-zoom},
but the precision of photometric redshifts
does not allow definitive answers to these kind of questions.

\subsection{Selection for Follow-up}

SDSS~J0903+5028 was recognized as a possible gravitational lens during
routine testing of the spectroscopic outputs of the SDSS.  There are
two independent SDSS software pipelines developed for
classifying spectra and assigning redshifts: spectro1d (briefly 
described by Stoughton et al. 2002; for more detail, 
see SubbaRao et al. 2003), 
which uses both cross-correlation via 
Fourier transforms with a family of templates and emission line 
identification,  
and specBS \citep{Schlegel}, 
which uses $\chi^2$ template fits in wavelength space.  Significant
discrepancies in redshifts and/or classifications between the two were
examined by eye\footnote{This totaled only 1.7\% of all spectra in 
SDSS Data Release 1, of which roughly half
were of too low S/N  to yield a meaningful redshift, usually correctly
classified as ``unknown'' by both pipelines.}.  
One common type of discrepancy arises when light from superposed
objects falls within the $3\arcsec$ spectroscopic fiber, and the 
two pipelines make different choices about which object's redshift 
to report. In DR1, 
there were half a dozen galaxy-quasar superpositions at very
different redshifts identified in this way, including SDSS~J0903+5028.
None of the others appear to be lenses.
In the case of SDSS~J0903+5028, spectro1d returned a classification of 
Galaxy with a redshift of $z=0.388$ at 94\% confidence, while specBS 
returned a classification of Quasar with a redshift (albeit incorrect) 
of $z=1.788$. 

In addition to comparison of the two pipelines, the spectro1d pipeline 
also flags spectra which cross-correlate with two templates at substantially 
different redshifts at high confidence level. Such was the case with 
SDSS~J0903+5028: spectro1d reported a significant (80\% confidence) $z=3.6$ 
cross-correlation peak for this spectrum with a quasar template. 
The relative confidence levels of the galaxy and quasar peaks 
are in line with expectation, given that the galaxy flux through the 
$3\arcsec$ fiber is about twice that of the quasar (see below). 

One of the unusual features of this lens is the fact that 
the target was identified as a luminous red galaxy, not as a quasar:   
the lensing galaxy is brighter than the lensed quasar images (see 
Table~\ref{tab-image-mod}). This system would therefore not be included in 
many optical searches for lensed quasars, because the brighter component 
has galaxy rather than quasar colors and because it was identified 
as an extended rather than a point source. On the other hand, in 
surveys that extend to faint magnitudes, it is not completely surprising 
to find such objects. For example, for UV-excess selected 
quasars at $z < 2.5$, Kochanek (1991) 
estimated that in a few percent of three-image lenses, the lens galaxy 
flux will exceed that of the combined quasar light for surveys to 
$m= 21$; presumably this percentage is higher for multi-band 
surveys that include quasars to higher redshifts. Alternatively, 
the typical $r$-band flux for a spectroscopically targeted 
$z=3.6$ quasar in the SDSS is $r \sim 19.4$ (PSF mag), while the typical  
$r$-band flux from a targeted $z=0.38$ LRG is $r \sim 18.8$ (model 
mag). While this comparison is obviously biased by our target selection 
criteria, it is nevertheless suggestive.

\subsection{ARC Spectrum}

We conducted follow-up spectroscopy of SDSS~J0903+5028 and other
interesting lens candidates on 2002 October 9 with the Astrophysical
Research Consortium (ARC) 3.5m telescope at Apache Point, New Mexico
using the Double Imaging Spectrograph\footnote{DIS II
see http://www.apo.nmsu.edu/Instruments/DIS/}. This instrument has a
dichroic at 5550\AA; the red and blue spectra combined have  
a usable wavelength coverage of about 3700\AA ~to 10000\AA.  We
took a 22 minute spectrum with the slit aligned along the direction
connecting the two primary image (A and B/G) components.  
Although the two spectra were partially 
blended and of relatively low signal-to-noise ratio, it was clear
after reduction that both spectra contained flux from 
high redshift quasars at the same redshift of $z\sim 3.6$.  
Both spectra also showed absorption features from the galaxy. 
Because the DIS data established a strong case for lensing but were not 
definitive, we subsequently re-observed this lens candidate at the 
W. M. Keck Observatory.

\subsection{Keck Spectrum}

\begin{figure}[p]
\epsscale{1.0}
\plotone{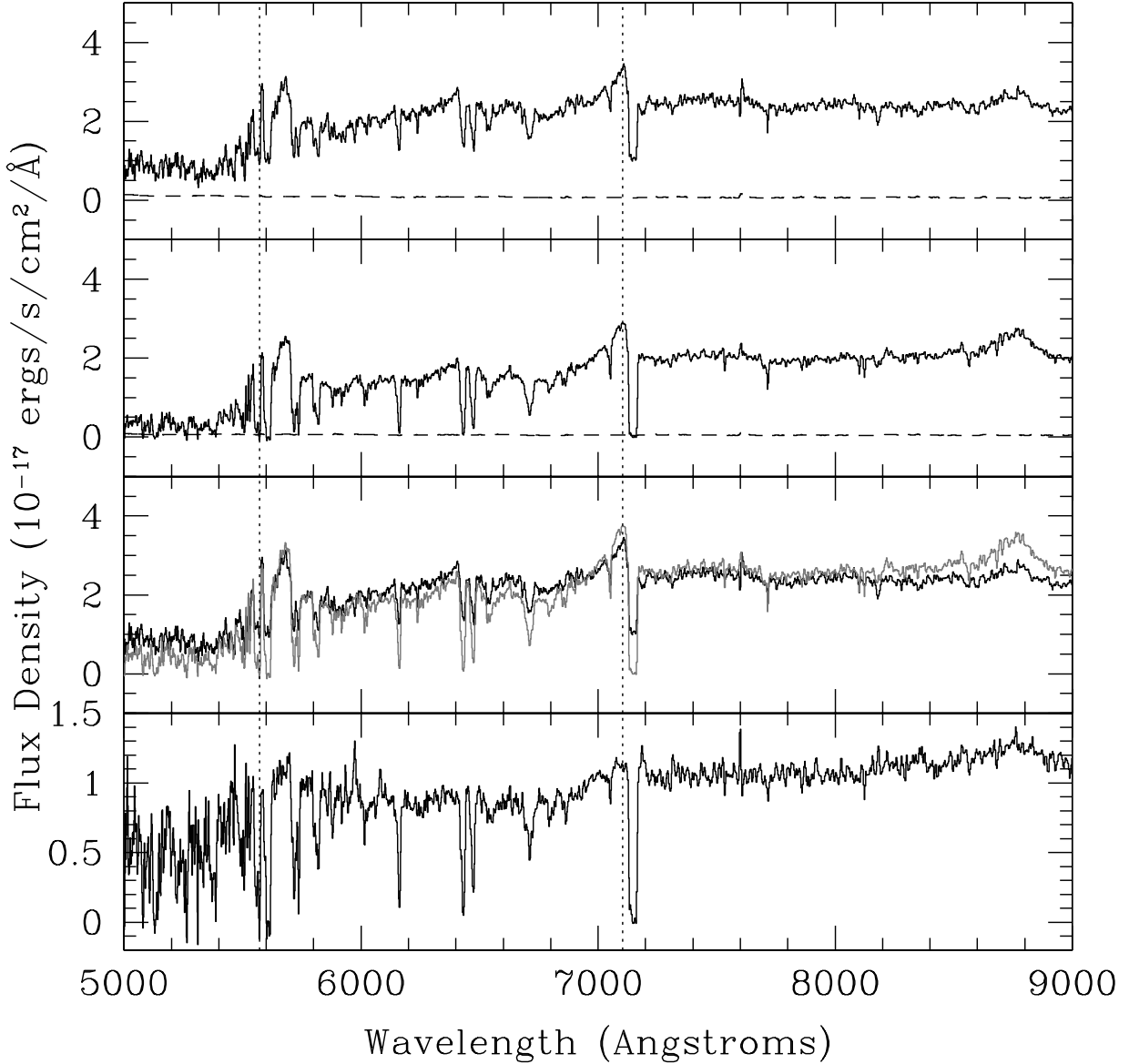}
\caption{Keck spectra of SDSS~J0903+5028. The dotted vertical lines 
show the location of Ly$\alpha$ and CIV at the peak CIV redshift 
of $z=3.584$. The best fit redshift is instead $z=3.605$. 
{\em Top:} Western (brighter `B/G') component.  
{\em Second:} Eastern (fainter `A`) component.  
{\em Third:} Fainter spectrum times 1.3 over-plotted on brighter component.  
{\em Bottom:} Ratio of scaled fainter component (A) to brighter (B/G) 
component. The larger contamination of the brighter component by 
the lensing galaxy makes the absorption line strengths appear 
different in the two spectra, when in fact they are very similar.
\label{fig:keckspec}}
\end{figure}

We obtained a high-dispersion
spectrum of both image components of SDSS~J0903+5028 using the echelle
spectrograph and imager (ESI; Epps \& Miller 1998) on the Keck II telescope
on the night of 2002 December 5; see Figure~\ref{fig:keckspec}.  Three
slit orientations were used; here we report only the pair of spectra
taken with the slit perpendicular to the axis separating the image pair, 
since these observations yielded the cleanest reductions. The night
was clear, with $0\farcs8$ seeing.  A 900s high-resolution spectrum
was taken for each member of the pair through a $1\arcsec$ slit 
in the echellette mode of ESI.  In this mode, the spectral range of
3900\AA ~to 11000\AA ~is covered in 10 spectral
orders with a nearly constant dispersion of $11.4\,{\rm km}\,{\rm
s^{-1}}\,{\rm pixel^{-1}}$.  Wavelength calibrations were performed
with observations of a CuAr lamp.  The spectrophotometric 
standard BD+28 4211 
was observed for flux calibration.  The data were reduced using a
tailored set of IRAF\footnote{IRAF is distributed by the National Optical
Astronomy Observatories, which are operated by the Association of
Universities for Research in Astronomy, Inc., under cooperative
agreement with the National Science Foundation.} and IDL 
routines developed specifically for
ESI data.  The smoothed spectra are shown in Figure~\ref{fig:keckspec}.
The signal-to-noise ratio is
$\sim5\,{\rm pixel}^{-1}$ in the raw spectra at 1450\AA ~in the rest-frame
and $\sim29\,{\rm pixel}^{-1}$ in the smoothed spectra.
As with the SDSS spectrum, the peak \ion{C}{4} redshift is 
$z=3.584$. However, given that there is associated absorption long-ward 
of the emission peak, this redshift is likely to be an underestimate.
Using templates from Richards et al. (2002), which allow for the 
possibility that \ion{C}{4} emission is blueshifted with respect 
to systemic and also allowing for reddening of the spectrum, we 
find a best-fit redshift of $z=3.605$ for the less contaminated 
'A' component, which would place the 
associated \ion{C}{4} absorption at roughly the systemic redshift 
(instead of being infalling).
Figure~\ref{fig:keckspec} shows that the two components have 
remarkably similar spectra and consistent redshifts. Scaling the 
fainter Eastern (`A') component by a factor 1.3 leads to a good match with
the brighter Western (`G/B') spectrum.
Furthermore the spectrum is by no means a typical quasar spectrum
since it has irregular BAL-like troughs. The fact that both components
have these same rare BAL troughs makes the lensing case very solid; 
the fact that there is clearly a galaxy between them makes the 
case practically certain.

\subsection{ARC Imaging}

From the SDSS imaging data, it was apparent that SDSS~J0903+5028 does not
simply comprise two point sources: the G/B source is
extended and was tentatively 
interpreted as a possible superposition of the lens galaxy with a
quasar point source. To further test the lensing hypothesis 
and to determine source positions and magnitudes for lens modeling, we 
obtained follow-up imaging data on 2002 November 13 with the ARC 3.5m
telescope using SPIcam.  SPIcam is a backside-illuminated 
SITe $2048\times2048$ CCD camera with $24\mu m$ pixels and a
plate scale of $0\farcs14\,{\rm pixel}^{-1}$, giving a field of view of
$4\farcm78$.  Because of the small pixels, this camera can take
advantage of very good seeing.  As it turned out, the seeing was
$1\farcs1$, a significant improvement over the SDSS $1\farcs5$. 
Also, the longer exposure, co-added SPIcam images are about 1.8 magnitudes deeper in
$r$ than the corresponding SDSS image.  
We obtained four dithers in each of the SDSS $r$ and $i$ bands for a
total exposure time of 20 minutes in each band.  These images were
de-biased, overscan-corrected, and flat-fielded in the usual manner
with IRAF. Figure \ref{fig:spicam-zoom} shows a $35\arcsec$ 
by $35\arcsec$ region around
SDSS~J0903+5028 from the co-added $r$ band SPIcam image.  The 
small group of galaxies is evident.  Based on modeling (see below), 
the objects labeled A and B were identified as 
the quasar images, while the object labeled G is the galaxy image, 
blended with quasar image B.
We used SExtractor \cite{sex} to find objects in 
the co-added $r$ and $i$ images and
matched these to the SDSS imaging catalog to obtain photometric
zero-points and an accurate astrometric solution.

\section{Analysis}

\subsection{Modeling the Image}

In order to fit a lens model to the data we proceed to determine the
positions and relative fluxes of the quasar and galaxy images.  While
ideally one would like higher resolution images for this purpose, 
we can in fact determine the configuration of this system 
quite confidently with
just arc-second imaging. The co-added ARC 3.5m images in both 
SDSS $r$ and $i$ filters are used to fit for an image model.

\begin{figure}[p]
\epsscale{1.0}
\plotone{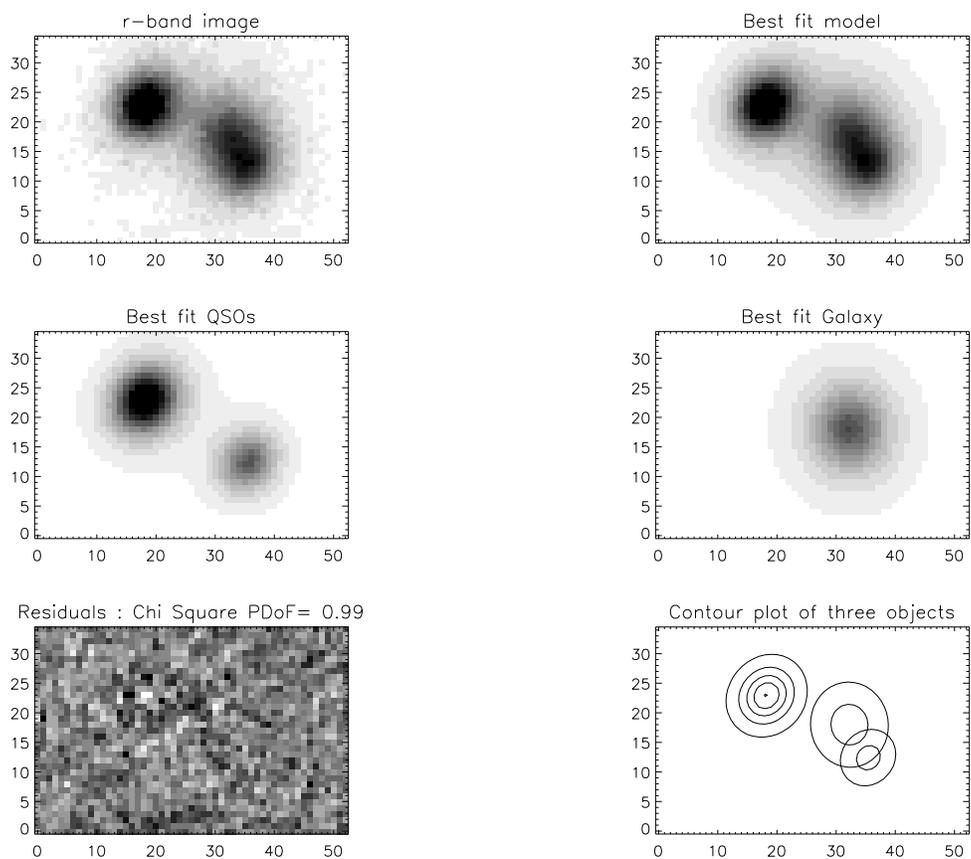}
\caption{Model fits for the SPIcam $r$-band image of 
SDSS~J0903+5028. {\em Top left:} Image. {\em Top right:} best fit model 
to two point sources and one extended source. {\em Middle:} Best fit 
model quasar and galaxy surface brightnesses. {\em Lower left:} Residuals 
between best fit model and the image. {\em Lower right:} Surface brightness 
contours for the 3 model components.
\label{fig:lens-model-r}}
\end{figure}

\begin{figure}[p]
\epsscale{1.0}
\plotone{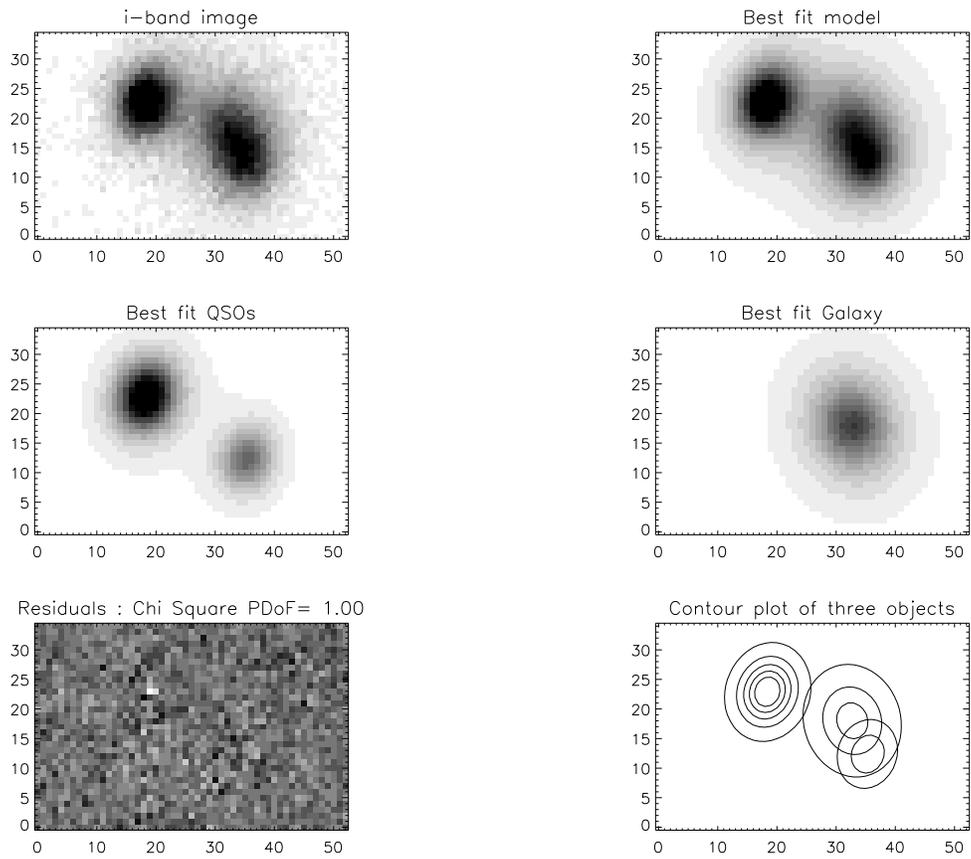}
\caption{Model fits for the SPIcam $i$-band image of SDSS~J0903+5028. 
For legend, see Figure~\ref{fig:lens-model-r}.
\label{fig:lens-model-i}}
\end{figure}

The top left panels in Figures \ref{fig:lens-model-r} and
\ref{fig:lens-model-i} show the $r$ and $i$ band co-added SPIcam
images of the $8\arcsec$ by $5\arcsec$ area around the lens. The
object on the left (East, component A) is unresolved and is one of
the quasar images. The object on the right (West) is resolved 
and it is evident from visual inspection that this
is in fact bimodal, with a point source, the quasar, to
the lower right (southwest) of the blended object centroid. 

This hypothesis can be tested
by fitting the image to a simple parametric model and looking at the
residuals.  The simplest model consists of a two-image lens with the
galaxy in between the two quasar images.  There are some conditions
that must be met for this image to be consistent with gravitational
lensing.  The quasars should have identical shapes, consistent with the 
local point-spread function (PSF), while 
the galaxy may be more extended. The three objects
should have positions in the two bands that are statistically consistent, and the
quasars should have nearly identical flux ratios. There are other
conditions that relate the flux ratios and the three image positions that
arise from the gravitational lens model; we address those
additional constraints in the Section 3.3.  

We fit the surface brightness of all three
objects as two-dimensional $t$-distributions, also known as Moffat profiles.
A normalized, 2D $t$-distribution is given by 
\[ \phi ({\bf x}) = \frac{1}{2 \pi} 
~~ |\Sigma|^{-1/2} (1+\delta / \nu) ^{-(\nu + 2)/2} ~, \]
where $ \delta = ({\bf x}- {\bf \mu})^T \Sigma^{-1} ({\bf x}- {\bf \mu}) $, 
the vector ${\bf \mu}$ is the image centroid, and $\Sigma$ is the $2 \times 2$
symmetric matrix of moments which determines the shape of
the elliptical isophotes.  The free parameter $\nu$ determines the
logarithmic slope of the asymptotic profile. In 
the limit $\nu \rightarrow \infty $, the surface brightness becomes a
Gaussian,  
$\phi({\bf x}) \rightarrow (2
\pi)^{-1} |\Sigma|^{-1/2} \exp(-\delta /2) $; even for finite 
$\nu$, $\phi$ is approximately Gaussian near the centroid.

Our PSFs are well fit by $\nu = 2$, so we fix $\nu$ to this value for the 
two point sources. 
The galaxy is also reasonably well fit by a $\nu = 2$ profile (but with 
different moments), but it is better fit by a $\nu = 1$ profile, so we
use the latter.  We further require that the quasars have the same
moments. This leaves 15 free parameters: 3 pairs of centroid coordinates, 3
fluxes, 3 PSF moments, and 3 galaxy moments. The fits are done independently
in each band. The best fit values are presented in Table \ref{tab-image-mod}.

The best fit dereddened magnitudes for the lens galaxy are 
$r=19.59 \pm 0.06$ and $i=18.86 \pm 0.07$, where we have included all
errors from shot noise, calibration, and model degeneracy. We can use the
measured redshift of 0.388 to calculate absolute magnitudes in both bands
and then use the $L-\sigma$ relation \citep{faber} to estimate the galaxy 
velocity dispersion in each band. We use $K$-corrections from 
Bruzual \& Charlot (1993) and correct for 
luminosity evolution using Bernardi et al. (2003) to arrive at
$M_r = -22.26$ and $M_i = -22.74$. The galaxy is therefore very luminous
;about $3L_*$ in both bands. Using the $L-\sigma$ relations and the 
$L-\sigma$ scatter from Bernardi et al. (2003), 
we estimate the velocity dispersion
as $\sigma_r = 206 \pm 53$ km/s and $\sigma_i = 213 \pm 54$ km/s. Typical lens
galaxy velocity dispersions are 200-300 km/s, so these values are 
not unusual; they are also consistent with the velocity dispersion inferred  
from the lens model below.  

The top two panels of Figures \ref{fig:lens-model-r} and
\ref{fig:lens-model-i} show the SPIcam data along with the best
fit model. The middle panels show the best fit model separated into the
quasars and galaxy. The lower panels show the residual image (image$-$model)  
and a contour
plot with the relative positions and fluxes of the three components. 
One can see that the $r$ and $i$ data give visually consistent results.  
The best fit models have
reduced $\chi ^2$ of 0.99 in $r$ and 1.00 in $i$, indicating that the
model is a good fit to both bands. The inferred quasar flux ratios (B/A) 
in $r$ and $i$ are $0.483 \pm 0.012$ and $0.461 \pm
0.021$, consistent at the 1.3 $\sigma$ level.
The quasar separations are 2\farcs83 $\pm$ 0\farcs02 in $r$ 
and 2\farcs80 $\pm$ 0\farcs03 in $i$, consistent at about 
the 1 $\sigma$ level. 

The image model also yields a measurement of the ellipticity of the 
galaxy light. The uncorrected model galaxy ellipticity is $\epsilon  
\equiv (1-r^2)/(1+r^2) =0.12$ 
in the $r$ image and 0.19 in $i$, with position angles 
of 12.0 and 18.2 deg (East of the North-South axis) 
in the two bands; here, $r =b/a$ is the ratio of semi-minor to 
semi-major axis of the surface brightness distribution. However, these
numbers do not take into account the extent and anisotropy of the image
PSF. We correct the galaxy shape measurement by subtracting 
the second moments of the local PSF from 
the second moments of the G model image. Using these deconvolved 
moments, the estimated corrected galaxy ellipticity is  
$\epsilon = 0.27$ in $r$ and $0.32$ in $i$; the 
corresponding position angles are 24.8 and 30.8 deg. The 
estimated error on the inferred ellipticity is about 0.1.

While the image modeling above does not rule
out more complicated lens configurations, it does show that this image
is consistent with the simplest configuration of a two-image lens.
We also note that we have applied this image 
fitting procedure to the lower signal-to-noise ratio SDSS images  
and to subsequent CFHT images (taken in better seeing but 
with more complex PSF structure) with very similar results.

\subsection{Modeling the Spectrum}

As with the imaging, we have also attempted to model the various spectra
of SDSS~J0903+5028 as a sum of quasar and galaxy components, using   
Principle Component Analysis (PCA). A large number of redshifted SDSS quasar  
spectra are used to construct eigenspectra $e_i(\lambda)$, which form an
orthonormal  basis in terms of which any other quasar spectrum can be expanded, 
$f_{\rm QSO}(\lambda) = \Sigma_i^N c_i e_i(\lambda)$. Similarly, a set of
galaxy  eigenspectra are constructed from many SDSS galaxy spectra. 
Spectra can be usefully classified by their 
coefficients $c_i$, provided they can be accurately  reconstructed when 
the series is truncated at relatively small $N$. Three 
eigenspectra span the range of most galaxy types; for quasars, more 
components are needed. Here, we used 10 quasar and 10 galaxy eigenspectra, 
constructed from samples of several thousand SDSS quasar spectra 
and about 100,000 SDSS galaxy spectra \citep{Yip}.  (The same galaxy
eigenspectra are used in the SDSS spectro1d spectroscopic pipeline
to classify galaxies.) 

\begin{figure}[p]
\epsscale{1.0}
\plotone{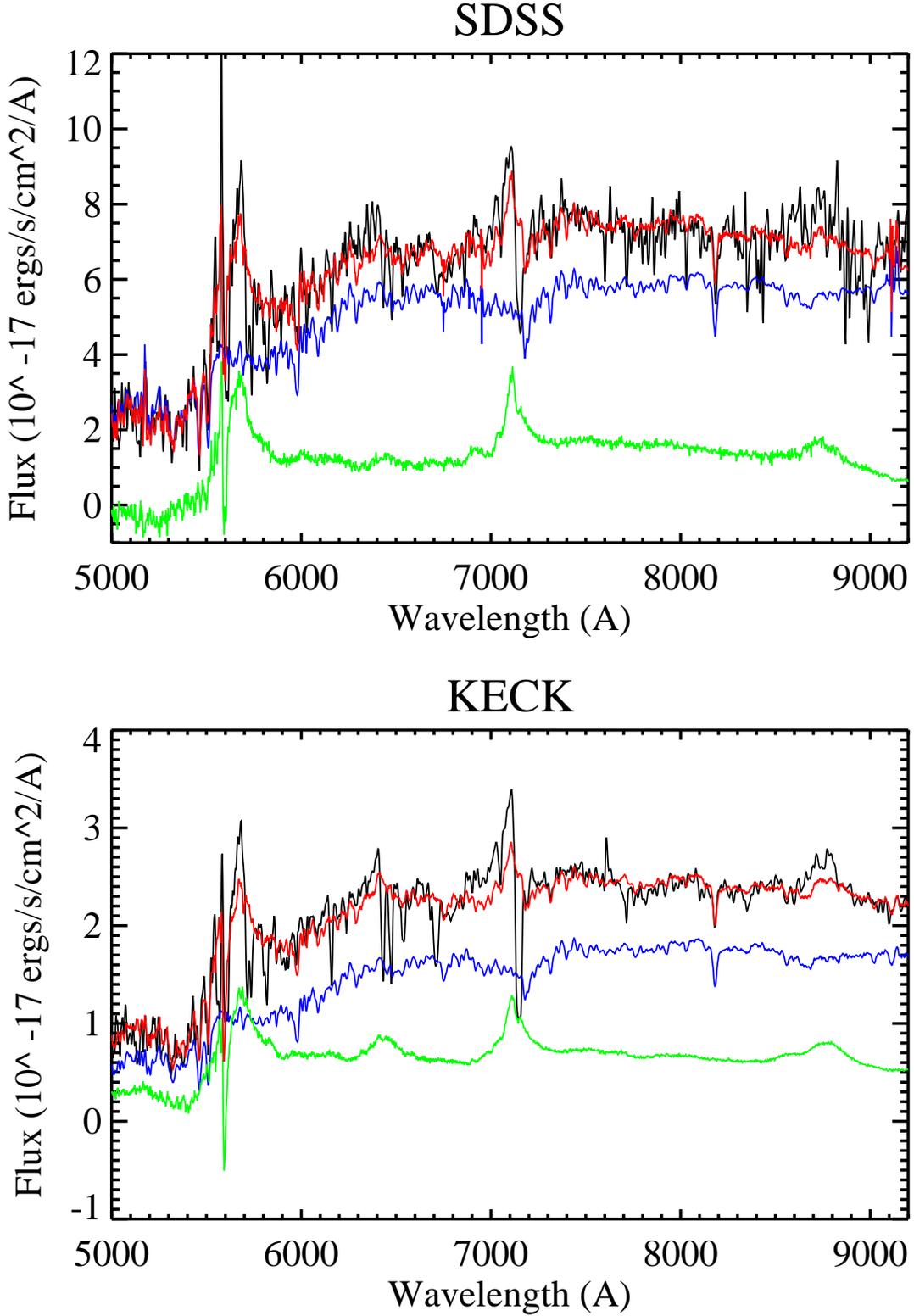}
\caption{{\em Top:} SDSS spectrum of SDSS~J0903+5028, 
decomposed into galaxy and quasar 
components: Spectrum (black), Galaxy (blue), quasar (green),  
sum of Galaxy and quasar components (red). {\em Bottom:} 
Keck Western spectrum of SDSS~J0903+5028, decomposed into 
LRG and quasar components. \label{fig:PCA}}
\end{figure}

To decompose a spectrum containing 
both quasar and galaxy components, for which the two redshifts are known, 
we simply assume it can be modeled as a weighted 
sum of the 
galaxy and quasar eigenspectra, where the coefficients are determined 
by minimizing the $\chi^2$ of the reconstructed spectrum fit to the true
spectrum. 
An example of this 20-parameter fit (hereafter called Model I) 
is shown in the top panel of Figure~\ref{fig:PCA}, 
which shows the best fit to the SDSS spectrum of the G/B component; for this 
fit, the quasar flux is about 25\% of the galaxy flux summed over this wavelength
range. Unfortunately, given the nature of this procedure, 
it is difficult to assign an error to this value. 

In addition to the model above, 
we experimented with two other models with fewer parameters. In 
Model II,  instead of 
using 10 galaxy eigenspectra, we fixed the galaxy spectrum to have the shape
of the average Luminous Red Galaxy spectrum  
constructed by co-adding a large number of LRG spectra \citep{eis2}. 
This makes 
use of the information that LRG spectra are quite homogeneous and that the 
G component has colors typical of an LRG. An example is shown 
in the lower panel of Figure~\ref{fig:PCA}, which shows a decomposition of
the Keck 
Western (G/B) spectrum using this model. As with the SDSS spectrum, the 
reconstructed quasar spectrum is a reasonable first approximation to the 
observed spectrum; not surprisingly, this procedure does not capture the 
BAL-like features, since the parent sample of SDSS quasar spectra used to 
produce the quasar eigenspectra did not include quasars with BAL features. 
The ratio of quasar to 
galaxy flux for this model is 42\%; for comparison, 
Model I for this spectrum yields a quasar/galaxy flux ratio of 48\% and yields 
a galaxy spectrum with the general spectral shape of an LRG. The differences  
between this spectral decomposition and that for the SDSS spectrum are 
not particularly  
troubling: the Keck spectrum has higher signal-to-noise ratio, and it is based on 
a narrow slit with rather different aperture from the SDSS fiber spectrum.  
On the other hand, the quasar/galaxy flux ratio for the Keck West spectrum model 
is in good agreement with that inferred 
from the $r$ and $i$ band imaging given in Table ~\ref{tab-image-mod}. 

In Model III, on the assumption that the Keck East spectrum 
has little contamination by the lensing galaxy, we fit 
the Keck West spectrum to a sum of Keck East and the LRG template or to 
Keck East plus 10 galaxy eigenspectra. This model generally gave poor 
or unphysical fits, consistent with the fact that 
Figure~\ref{fig:keckspec} appears to indicate that the Western component  
is somewhat bluer than the Eastern component. 
The latter result is somewhat surprising: given the lens geometry 
shown in Figure~\ref{fig:lens-model-r} and the results in 
Table~\ref{tab-image-mod}, one would naively expect the 
Eastern component to be less contaminated by the red lensing 
galaxy than the Western component, assuming the East and West 
spectral components correspond approximately to the A and B/G image components 
of Figure~\ref{fig:spicam-zoom}. This apparent discrepancy may be due 
in part to the spectral extraction algorithm, errors in relative 
spectrophotometric calibration, placement of the slit,  
intrinsic reddening of the quasar spectra, differential reddening 
in the galaxy, or quasar spectral variability 
on a timescale shorter than the time delay between the images. 

Finally, we also attempted to measure the galaxy velocity dispersion 
from the 
quasar-subtracted galaxy spectrum, but it was too contaminated
by residual quasar absorption features to obtain a reliable result.

\subsection{Modeling the Lens}

To extract physical properties of the lens galaxy and its 
environment and to further test the lens hypothesis, we proceed 
to make lens models using the astrometry and photometry from the
model analysis of the images.  The uncertainties on the relative
positions are $0\farcs02$ in $r$ and $0\farcs03$ in $i$.  For the
fluxes, we broaden the error bars to 10\% to account for variability,
microlensing, etc.\ (see Dalal \& Kochanek 2002).
We use standard isothermal lens models, because they are
consistent with the observed properties of other individual lenses,
lens statistics, and the dynamics and X-ray properties of elliptical
galaxies \citep{fabbiano,kochanek1,kochanek2,maoz,rix,treu,rusin}.
For the modeling we use $\Omega_m$ = 0.3 and $\Omega_\Lambda$ = 0.7,
although these only affect the reported velocity dispersions and 
time delays, and then only at the few percent level.
We use standard non-linear least-squares lens modeling techniques,
implemented in the {\it lensmodel\/} software by Keeton (2001b). 
The data provide eight constraints: two each for two quasar image
positions, two for the galaxy position, and two fluxes.  A minimal
model has eight parameters: the galaxy position (2) and mass (1), the
ellipticity of the mass distribution or alternatively 
shear and its orientation angle (2), and the quasar source
position (2) and flux (1).  Even minimal models therefore have
$N_{\rm dof}=0$, and so we are always able to find models that fit
the data perfectly.  Hence to estimate the uncertainties on the
model parameters, we repeatedly add random noise to the 8 data points and
refit to obtain a distribution of fitted parameter values.

The fact that the quasar images and the galaxy are not collinear
indicates a non-negligible quadrupole moment in the lensing potential,
which may represent ellipticity in the lens galaxy and/or tidal shear
from mass in the environment of the galaxy \citep{kks}.  The presence
of mass ellipticity might be expected because the deconvolved galaxy
light is elliptical; moreover, the mass could be more flattened than
the light.  The presence of shear seems likely because of the
surrounding galaxy clusters.  To consider both possibilities, we first
examine two simple models:
(1) a singular isothermal ellipsoid (SIE) model, where the quadrupole
is due entirely to ellipticity, and
(2) a singular isothermal sphere (SIS) plus shear model, where the
quadrupole is due entirely to tidal shear.

Both SIE and SIS+shear models can fit the lens exactly, with the
parameters given in Table~\ref{tab-mod}.  Both sets of models seem
reasonable: SIE models require a mass ellipticity $\epsilon=0.5$--0.6,
slightly larger than the ellipticity of the light ($\sim\!0.3$), 
while SIE+shear models require a shear strength
$\gamma=0.15$--0.18, typical of lenses in group or cluster
environments \citep{kks,khb,kch}.  There are small differences between
the $r$-band and $i$-band models due to differences in the deconvolved
positions of quasar B and the galaxy in the $r$ and $i$-band data, but the
differences are only at the 1$\sigma$ level.

The models yield an Einstein radius of $1\farcs4$, corresponding to
a velocity dispersion of $250\pm 4$ ~km~s$^{-1}$ for the lens galaxy.  
This number is consistent with the velocity dispersion
estimates made with the $L-\sigma$ relations in Section 3.1; the 
estimate from the lens model may be higher due to the surrounding group 
slightly enhancing the image angular separation. 
The implied total magnification of the system is a moderate factor
of 3--4.  The models also predict that the time delay between the
images should be in the range 57--72~$h^{-1}$~days.  Because the
predicted delay depends on the relative amounts of ellipticity and
shear \citep{witt}, the usefulness of this lens for Hubble parameter
analyses will depend on how well the ellipticity and shear can be
determined independently.

\begin{figure}[p]
\plotone{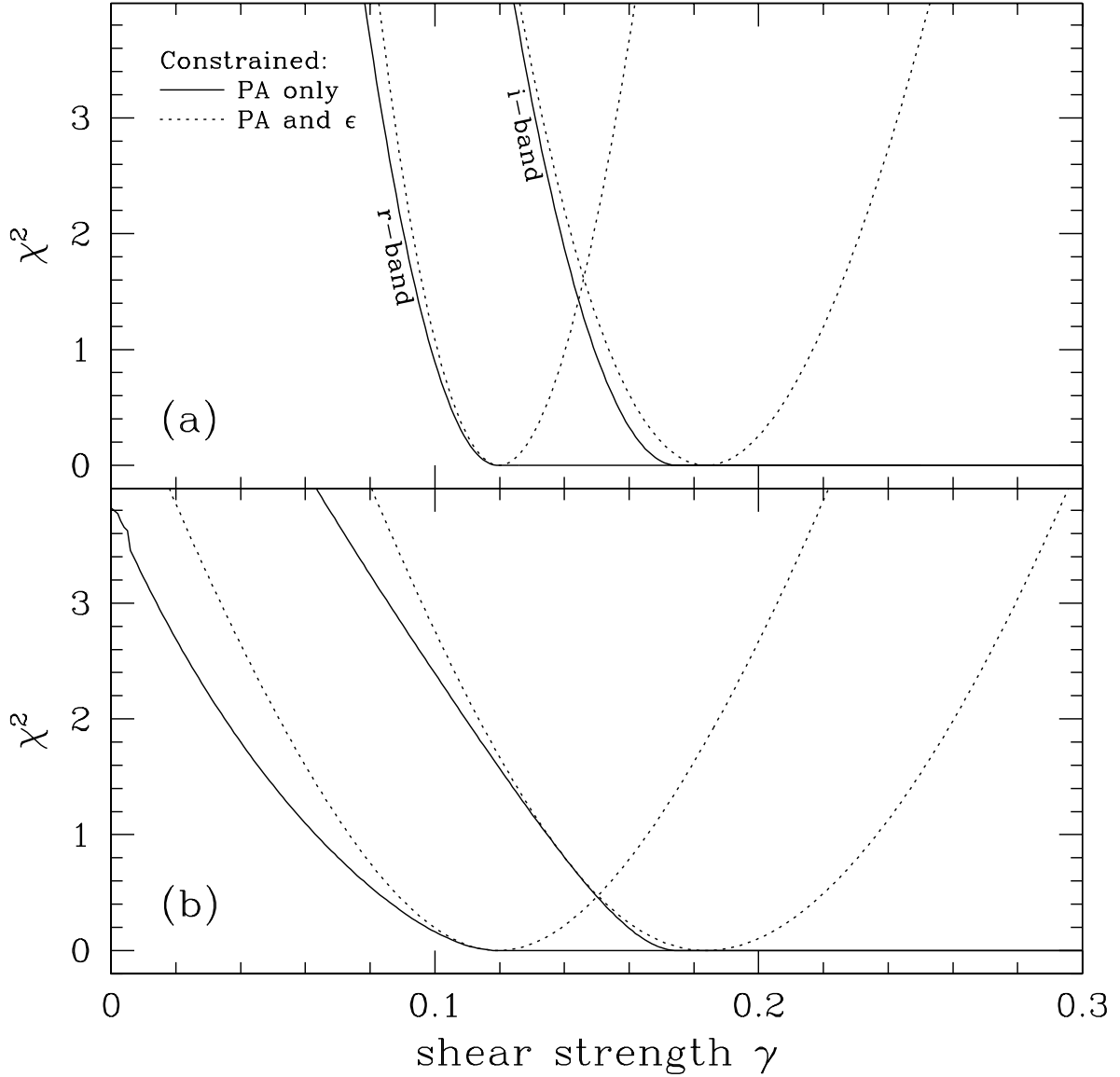}
\caption{
Results from SIE+shear lens models.  The solid curves show models in which the
position angle (PA) of the mass is constrained to match that of the light,
while the dotted curves show models for which both the position angle and
ellipticity of the mass are constrained to match those of the light.
There are separate curves for models of the $r$ and $i$-band data. 
(a) The mass, PA, and ellipticity are fixed.
(b) The mass, PA, and ellipticity are free parameters but constrained by
the light, with assumed uncertainties of $10^\circ$ and 0.1, respectively.
Note that the $\chi^2_{min}=0$ only because
$N_{\rm dof}=0$.\label{fig:shear}}
\end{figure}

It is interesting to note that in both SIE and SIS+shear models
the quadrupole moment of the lensing potential is
oriented almost exactly north--south, while the corrected galaxy light
is inclined at $\sim\!30^\circ$.  A misalignment of more than
$\sim\!10^\circ$ usually indicates that the lensing potential has both
ellipticity and shear with different orientations \citep{kkf,csk02}.
It is pointless to fit models with unconstrained ellipticity and shear
to SDSS~J0903+5028, because such models are under-constrained.  However,
analyses of other lenses suggest that it is reasonable to constrain
the shape of the model mass distribution using the observed shape of
the light distribution.  The orientation angles of the mass and light
are strongly correlated and typically aligned to within $\sim\!10^\circ$,
even if there is no clear relationship between the ellipticities of
the mass and light \citep{kkf,csk02}.  Figure~\ref{fig:shear} shows
results for SIE+shear models where we either fix the shape of the model mass
distribution to that of the light (panel a) or just require that the mass
distribution match the light within assumed uncertainties of $10^\circ$
in orientation and 0.1 in ellipticity (panel b).  In both cases, the
constraint on the mass orientation provides an important lower limit
on the shear strength.  In other words, under the reasonable assumption
that the mass distribution is aligned with the light
distribution, the misalignment between the galaxy and the quadrupole
moment of the lensing potential directly implies the presence of shear
from the lens environment (the group or nearby clusters).  
Adopting a constraint on the
mass ellipticity would then yield an upper limit on the shear, but
this result is less reliable because there is no strong evidence that
the mass ellipticity should match that of the light \citep{kkf}.
We also note that if the two nearby clusters found by the 
maxBCG algorithm are indeed separate spherical clusters,
they would produce a shear with the requisite north--south orientation; 
however, given their relatively large angular separation 
from the lens galaxy, 
one would expect them to produce a combined shear of only a few
percent.

Thus, the lens models suggest but do not conclusively reveal that the
group or clusters around the lens galaxy play an important role in the lensing
potential.  The best way to test this hypothesis would be to obtain
spectroscopy for galaxies in the field of the lens, to confirm the
cluster(s) and identify members, and to measure the centroid and velocity
dispersion of the cluster(s).  Those quantities could be used to estimate
the shear from the environment, and compared with the predicted shear
strength $\gamma \sim 0.1$--0.2 and orientation
$\theta_\gamma=136^\circ$--$174^\circ$ to test and further constrain
the lens models.

\section{Conclusions}

We have identified a lensed quasar candidate, SDSS~J0903+5028, 
based on the superposition of a $z=3.605$ quasar and a $z=0.388$ luminous red 
galaxy in an SDSS spectrum. Follow-up observations with the 
ARC 3.5-m and the Keck II telescope have confirmed that this is 
a two-image gravitational lens system, with image angular separation of 
2\farcs8. The lens model is consistent with a massive galaxy with a
velocity dispersion of 250 km sec$^{-1}$. The lens geometry indicates
a quadrupolar lensing potential which can be generated by an elliptical
galaxy mass distribution and/or tidal shear from what appears to be
a group of galaxies surrounding the lens.  The misalignment between
the quadrupole and the galaxy light suggests that there is indeed
significant shear from the environment.

\section{Acknowledgments}

We thank Paul Schechter and Scott Burles 
for useful discussions. 
Funding for the creation and distribution of the SDSS Archive has been
provided by the Alfred P. Sloan Foundation, the Participating
Institutions, the National Aeronautics and Space Administration, the
National Science Foundation, the U.S. Department of Energy, the
Japanese Monbukagakusho, and the Max Planck Society. The SDSS Web site
is http://www.sdss.org/. The SDSS is managed by the Astrophysical
Research Consortium (ARC) for the Participating Institutions. The
Participating Institutions are The University of Chicago, Fermilab,
the Institute for Advanced Study, the Japan Participation Group, The
Johns Hopkins University, Los Alamos National Laboratory, the
Max-Planck-Institute for Astronomy (MPIA), the Max-Planck-Institute
for Astrophysics (MPA), New Mexico State University, Princeton
University, the United States Naval Observatory, the University of 
Pittsburgh, and the University of Washington. JF and DJ acknowledge 
support from the NSF Center for Cosmological Physics and NSF 
grant PHY-0079251, from the DOE, and from NASA grant NAG5-10842. 
GTR acknowledges support from HST-GO-09472.01-A. Part of the work 
reported here was done at LLNL under the auspices of the U.S. 
Department of Energy under contract W-7405-Eng-48. This work is 
based in part on observations obtained with the Apache Point Observatory 
3.5-meter telescope, which is owned and operated by the 
Astrophysical Research Consortium. Some of the data presented 
herein were obtained at the W.M. Keck Observatory, which is operated 
as a scientific partnership among the California Institute of Technology, 
the University of California, and the National Aeronautics and 
Space Administration. The Observatory was made possible by the generous 
financial support of the W.M. Keck Foundation. 
We thank the staffs of
the Keck and Apache Point Observatories, and 
C. Ryan at CFHT, for their assistance.

\clearpage

\clearpage

\begin{deluxetable}{ccccccc}
\tablecaption{SDSS Photometry}
\tablewidth{0pt}
\tablehead{
 \colhead{Object} &
\colhead{Mag} &
 \colhead{$g$} &
 \colhead{$r$} &
 \colhead{$i$} &
\colhead{$z$}
}
\startdata
A & Petro      
&$21.20\pm 0.18  $ & $19.53\pm 0.12 $ & $19.16\pm0.11  $ & $19.14\pm 0.34 $   
\\
& PSF &
$21.78\pm 0.08$ & $20.22\pm 0.06$ & $19.58\pm 0.04$ & $19.27\pm 0.09$
\\
\tableline 
B/G & Petro  
& $20.85\pm0.39$ & $19.28\pm 0.14$ & $18.73\pm 0.18$ & $18.38\pm 0.22$  \\
& Model & 
$21.33\pm 0.10 $ & $19.26\pm 0.02$ & $18.50\pm 0.02$ & $18.17\pm 0.04$
\\
\enddata
\tablecomments{$griz$ Petrosian (1976), PSF, and model  
magnitudes returned by SDSS photometric pipeline reduction rerun 21 for 
run 2074, camcol 2, field 113, object id 185 (A) and 186 (B/G). 
PSF magnitudes are appropriate for point sources (component A, to 
the extent it is deblended), 
while model magnitudes \citep{Stoughton} are used to 
define colors for LRG targeting \citep{eis} (component B/G). 
The model magnitude errors only include residuals from the 
model profile fit and are therefore artificially low.
These are asinh magnitudes \citep{lgs99} 
and are corrected for Galactic reddening
according to the dust map of Schlegel, Finkbeiner, \& Davis (1998). 
$u$ magnitudes are not reported since no significant flux was measured 
in this band. More accurate  
photometry based on deeper ARC 3.5m imaging is presented in 
Table~\ref{tab-image-mod} below.
}
\label{tab-image-sdssmag}
\end{deluxetable}

\begin{deluxetable}{ccccccc}
\tablecaption{ARC Photometry: Model Results}
\tablewidth{0pt}
\tablehead{
 \colhead{Object} &
 \colhead{(J2000)} &
 \colhead{$r$} &
 \colhead{$i$} &
 \colhead{ $r-i$}
}
\startdata  
QSO A & 
$09^{\rm h} 03^{\rm m} 35^{\rm s}.132+50^{\circ} 28\arcmin 20\farcs21$
&$19.99 \pm 0.01$ & $19.43 \pm 0.01$ & $0.56 \pm 0.02$   \\
QSO B       & 
$09^{\rm h} 03^{\rm m} 34^{\rm s}.877+50^{\circ} 28\arcmin 18\farcs75$
& $20.78 \pm 0.03$ & $20.27 \pm 0.05$ & $0.51 \pm 0.06$  \\
Galaxy       & 
$09^{\rm h} 03^{\rm m} 34^{\rm s}.925+50^{\circ} 28\arcmin 19\farcs53$
& $19.59 \pm 0.02$ & $18.86 \pm 0.04$ & $0.73 \pm 0.05$   \\
\enddata
\tablecomments{Dereddened magnitudes. 
The astrometry is well calibrated to the SDSS astrometry
and so the dominant errors simply come from the fitting routine
and this error is about 0\farcs07. The photometry is also calibrated
to the SDSS and has a systematic error in the zero points estimated
at 0.06 in both bands. The relative photometry could in principle
be better but, due to the degeneracy between QSO B and the galaxy,
the error on the relative magnitudes are at about the same level. 
We conclude that the color difference between the two quasars 
is consistent with zero.
}
\label{tab-image-mod}
\end{deluxetable}

\begin{deluxetable}{ccccccc}
\tablecaption{MaxBCG Clusters}
\tablewidth{0pt}
\tablehead{
 \colhead{(J2000)} &
 \colhead{angle} &
 \colhead{Distance} &
 \colhead{$z$} &
 \colhead{$N_{gal}$} 
}
\startdata
$09^{\rm h} 03^{\rm m} 43^{\rm s}+50^{\circ} 32\arcmin 58\arcsec$
& $5\farcm7$ & 1.36 & 0.44 & 12 \\
$09^{\rm h} 03^{\rm m} 47^{\rm s}+50^{\circ} 24\arcmin 54\arcsec$
& $5\farcm8$ & 1.39 & 0.44 & 14 \\
\enddata
\tablecomments{The two clusters near the lens galaxy.
The first column gives the cluster center J2000 coordinates as reported 
by the maxBCG algorithm. The second column is the separation in
arc-minutes of the cluster center
from the lens galaxy. The third is the separation in
Mpc at the indicated redshift. The fourth column is the photometric
redshift as reported by maxBCG. The estimated errors on the maxBCG  
redshift estimates are
typically 0.02 for low redshift and about 0.05 for these higher
redshift clusters. The fifth column is  
$N_{gal}$, a richness measure returned by maxBCG, 
the estimated number of $L^*$ and brighter galaxies in the cluster.
}
\label{tab-maxbcg}
\end{deluxetable}

\begin{deluxetable}{ccccccc}
\tablecaption{Lens Model Results}
\tablewidth{0pt}
\tablehead{
 \colhead{Type} &
 \colhead{Band} &
 \colhead{$R_E$ (\arcsec)} &
 \colhead{$\epsilon$ or $\gamma$} &
 \colhead{$\theta_\epsilon$ or $\theta_\gamma$ (\arcdeg)} &
 \colhead{$\mu_{\rm tot}$} &
 \colhead{$\Delta t$ ($h^{-1}$ days)}
}
\startdata
SIE       & $r$ & $1.42\pm0.03$ & $0.47\pm0.04$ & $4.6\pm2.8$ & $3.62\pm0.19$ & $67.4\pm2.8$ \\
          & $i$ & $1.38\pm0.04$ & $0.57\pm0.06$ & $2.0\pm3.0$ & $3.19\pm0.23$ & $72.0\pm4.2$ \\
\tableline
SIS+shear & $r$ & $1.44\pm0.03$ & $0.15\pm0.01$ & $3.1\pm3.0$ & $4.13\pm0.21$ & $57.0\pm2.2$ \\
          & $i$ & $1.43\pm0.04$ & $0.18\pm0.02$ & $1.0\pm3.2$ & $3.76\pm0.26$ & $57.8\pm3.1$ \\
\enddata
\tablecomments{
Col.\ 3 gives the Einstein radius.
Col.\ 4--5 give the ellipticity $\epsilon$ and position angle $\theta_\epsilon$
for SIE models, or the shear $\gamma$ and position angle $\theta_\gamma$ for
SIS+shear models.
Col.\ 6 gives the total magnification.
Col.\ 7 gives the predicted time delay (for a cosmology with
$\Omega_M=0.3$ and $\Omega_\Lambda=0.7$).
}
\label{tab-mod}
\end{deluxetable}

\end{document}